\documentclass[aps,pre,superscriptaddress,twocolumn,showpacs]{revtex4}

\usepackage{epsfig}
\usepackage{amsmath}
\usepackage{times}
\usepackage{graphicx}
\usepackage{amssymb}
\usepackage{multirow}

\begin{document}

\title{Convergence to global consensus in opinion dynamics under a nonlinear voter model}

\author{Han-Xin Yang}\email{hxyang01@gmail.com}
\affiliation{Department of Physics, Fuzhou University, Fuzhou
350002, P. R. China} \affiliation{Department of Modern Physics,
University of Science and Technology of China, Hefei 230026, China}

\author{Wen-Xu Wang}
\affiliation{School of Electrical, Computer and Energy Engineering, Arizona State
University, Tempe, AZ 85287}

\author{Ying-Cheng Lai}
\affiliation{School of Electrical, Computer and Energy Engineering,
Arizona State University, Tempe, AZ 85287} \affiliation{Department
of Physics, Arizona State University, Tempe, AZ 85287}

\author{Bing-Hong Wang}
\affiliation{Department of Modern Physics, University of Science and
Technology of China, Hefei 230026, China} \affiliation{Research
Center for Complex System Science, University of Shanghai for
Science and Technology and Shanghai Academy of System Science,
Shanghai, 200093 China}

\begin{abstract}

We propose a nonlinear voter model to study the emergence of global
consensus in opinion dynamics. In our model, agent $i$ agrees with
one of binary opinions with the probability that is a power function
of the number of agents holding this opinion among agent $i$ and its
nearest neighbors, where an adjustable parameter $\alpha$ controls
the effect of herd behavior on consensus. We find that there exists
an optimal value of $\alpha$ leading to the fastest consensus for
lattices, random graphs, small-world networks and scale-free
networks. Qualitative insights are obtained by examining the
spatiotemporal evolution of the opinion clusters.

\end{abstract}

\date{\today}


 \maketitle

\section{Introduction}

Mutual agreement, or consensus, is a fundamental phenomenon in
social and natural systems. The dynamics of opinion sharing and
competing and the emergence of consensus have become an active topic
of recent research in statistical and nonlinear physics
\cite{physics}. For example, a number of models have been proposed
to address how consensus can result from the evolution of two
competing opinions in a population, which include the voter model
\cite{voter1,voter2}, the majority rule model
\cite{majority1,majority2}, the bounded-confidence model
\cite{confidence} and the social impact model \cite{social1}. Due to
the high relevance of complex networks \cite{complex network} to
social and natural systems, opinion dynamics have also been
incorporated on networks
\cite{network0,network1,network3,network9,network12.1,
network12.2,network15,network16,network17.1,network18,other1,other2,other3,other4,other5,other6,other7,other8,other9,other9.1,other10,other11,other12,other12.1,other13,other14,other15,other16}
such as regular lattices, random graphs \cite{ER}, small-world
networks \cite{NW} and scale-free networks \cite{BA}.

Previous works have revealed phase transitions in opinion dynamics
\cite{network0,network1,network12.1,network15,network16} and the
emergence of global consensus \cite{voter2,majority2} where all
agents share the same opinion. It has also been found that both the
network structures \cite{network9} and the opinion updating
strategies \cite{network12.2,network17.1,network18} can affect the
time for reaching the final consensus. In Ref. \cite{yang1}, Yang
$et$ $al.$ combined the majority rule model with probability $p$
with the voter model with probability $1-p$ and then measure the
resulting consensus times on scale-free networks. They found that
that the optimized ratio to minimize consensus time is around
$p=0.72$. In Refs. \cite{yang2,yang3,yang4}, Yang $et$ $al.$ studied
the effects of heterogeneous influence of individuals on the global
consensus. Each individual is assigned a weight that is proportional
to the power of its degree, where the power exponent $\alpha$ is an
adjustable parameter. Interestingly, it is found that there exists
an optimal value of $\alpha$ leading to the shortest consensus time
for scale-free networks, random networks and small-world networks.

In the voter model, an agents follows the opinion of a randomly
selected neighboring agent. In the majority rule model, an agent
follows the local majority opinion. For the voter model and the
majority rule model, an agents absolutely follows one opinion.
However, in reality, there can be situations where an agent chooses
one opinion with a stochastic probability due to bounded
rationality. In this Letter, we propose a nonlinear voter model in
which an agent $i$ selects one of binary opinions with the
probability that is proportional to the power of the number of
agents carrying this opinion among agent $i$ and its nearest
neighbors. The power exponent $\alpha$ is introduced to control
selective probability. Our main finding is that, there exists an
optimal value of $\alpha$ for which the convergent time for
consensus is minimum. This phenomenon indicates that, a suitable
preference of the local majority opinion (not following it
absolutely) can greatly accelerate the convergence process towards
the final consensus.

\section{Model}

Our model is described as follows. For a given network of any
topology, each node represents an agent. Initially the two opinions
denoted by the values $\pm1$ are randomly assigned to agents with
equal probability. At each time step, agents synchronously update
\cite{note} their opinions according to the following rule. Agent
$i$ will select the opinion $+1$ with the probability
\begin{equation}
p_{+}=\frac{n_{+}^{\alpha}}{n_{+}^{\alpha}+n_{-}^{\alpha}},
\end{equation}
where $n_{+}$ and $n_{-}$ are the number of agents holding the
opinion $+1$ and $-1$ among agent $i$ and its nearest neighbors
respectively, and $\alpha$ is an adjustable parameter. Similarly,
agent $i$ selects the opinion $-1$ with the probability
$p_{-}=n_{-}^{\alpha}/(n_{+}^{\alpha}+n_{-}^{\alpha})$, where
$p_{-}=1-p_{+}$.

The parameter $\alpha$ characterizes the degree of herd effect on
consensus. For $\alpha>0 \ (<0)$, agent $i$ has a larger probability
to select the local majority (minority) opinion. In the case of
$\alpha=0$, agent $i$ randomly selects the opinion $+1$ or $-1$. For
$\alpha=1$ , our model coincides with the voter model. For
$\alpha\rightarrow\infty$, agent $i$ absolutely takes the local
majority opinion and our model is reduced to the model proposed in
Ref. \cite{network15}.

\section{Results and analysis}

We first consider a square lattice with periodic boundary
conditions. Following Ref. \cite{network1}, we define an order
parameter $\eta$ as
\begin{equation}
\eta=\frac{1}{N}\mid\sum_{i=1}^{N}\sigma_{i}\mid,
\end{equation}
where $\sigma_{i}$ is the value of node $i$'s opinion ($+1$ or $-1$)
and $N$ is the total number of nodes of the network. In general, we
have $0\leq\eta\leq1$, and a large value of $\eta$ indicates that
one opinion in the system dominates the other. If the two opinions
are equally probably, we have $\eta=0$. When global consensus is
achieved so that there is only one opinion in the system, we have
$\eta=1$.

\begin{figure}
\begin{center}
\scalebox{0.82}[0.82]{\includegraphics{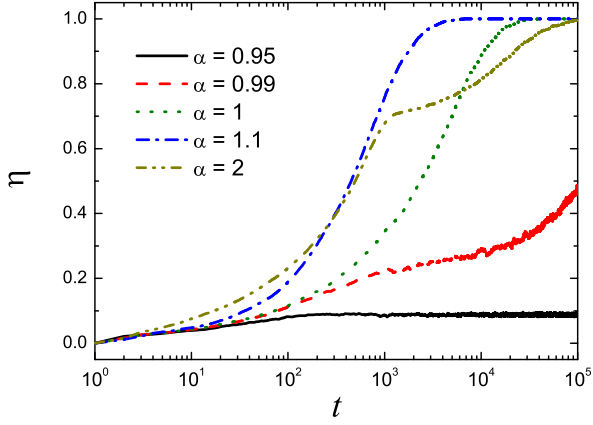}} \caption{(Color
online.) Order parameter $\eta$ as a function of $t$ for different
values of $\alpha$ on a $50\times50$ square lattice with periodic
boundary conditions. Each data point is obtained by averaging over
2000 different realizations.} \label{order}
\end{center}
\end{figure}

Figure~\ref{order} shows the order parameter $\eta$ as a function of
the time step $t$ for different values of $\alpha$. For $\alpha$=1,
1.1 and 2, $\eta$ increases from 0 to 1 with $t$. For $\alpha=2.0$,
initially $\eta$ increases faster than the cases of $\alpha=1.0$ and
$\alpha=1.1$. Nevertheless, the system converges to the global
consensus state faster for $\alpha=1.1$ than for $\alpha=2.0$.
Another feature in Fig.~\ref{order} is that, for $\alpha<1$, the
time required to converge to global consensus can be extremely long.

\begin{figure}
\begin{center}
\scalebox{0.4}[0.4]{\includegraphics{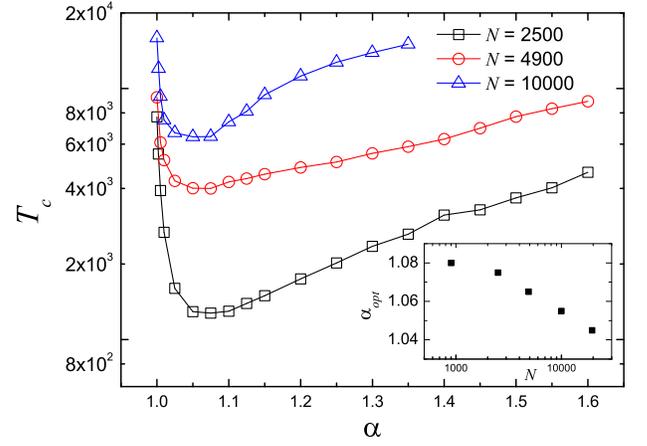}} \caption{Time to
achieve consensus, $T_{c}$, as a function of $\alpha$ for different
lattice size $N$. The inset shows the optimal value of $\alpha$,
 $\alpha_{opt}$, as a function of $N$. Each data point is obtained by
averaging over 2000 different realizations.} \label{lattice}
\end{center}
\end{figure}

Why does the system become so difficult to reach consensus for
$\alpha<1$? According to the mean-field approximation, opinion
dynamics in the preferential-selection model can be described as:
\begin{gather}
\frac{d\rho_{+}}{dt}=\frac{\rho_{+}^{\alpha}\rho_{-}}{\rho_{+}^{\alpha}+\rho_{-}^{\alpha}}-\frac{\rho_{-}^{\alpha}\rho_{+}}{\rho_{+}^{\alpha}+\rho_{-}^{\alpha}}\notag\\=\frac{\rho_{+}^{\alpha}(1-\rho_{+})}{\rho_{+}^{\alpha}+(1-\rho_{+})^{\alpha}}-\frac{\rho_{+}(1-\rho_{+})^{\alpha}}{\rho_{+}^{\alpha}+(1-\rho_{+})^{\alpha}}\notag\\
=\frac{\rho_{+}(1-\rho_{+})}{\rho_{+}^{\alpha}+(1-\rho_{+})^{\alpha}}[\rho_{+}^{\alpha-1}-(1-\rho_{+})^{\alpha-1}],
\end{gather}
where where $\rho_{+}$ and $\rho_{-}$ are the fractions of agents
holding the opinion $+1$ and $-1$ in the total population,
respectively. In the case of $\alpha<1$, for $\rho_{+}>0.5\ (<0.5)$,
$d\rho_{+}/dt<0\ (>0)$. This indicates a negative feedback mechanism
that prohibits the majority opinion from spreading over the whole
population.

We define the consensus time $T_{c}$ as the time steps required to
reach the final consensus. Figure~\ref{lattice} shows that $T_{c}$
as a function of $\alpha$ for different network size $N$. It can be
seen that there exists an optimal value of $\alpha$, hereafter
denoted by $\alpha_{opt}$, which results in the shortest consensus
time $T_{c}$. We have found that $\alpha_{opt}$ decreases as the
network size $N$ increases, as shown in the inset of
Fig.~\ref{lattice}.

\begin{figure}
\begin{center}
\scalebox{0.96}[0.96]{\includegraphics{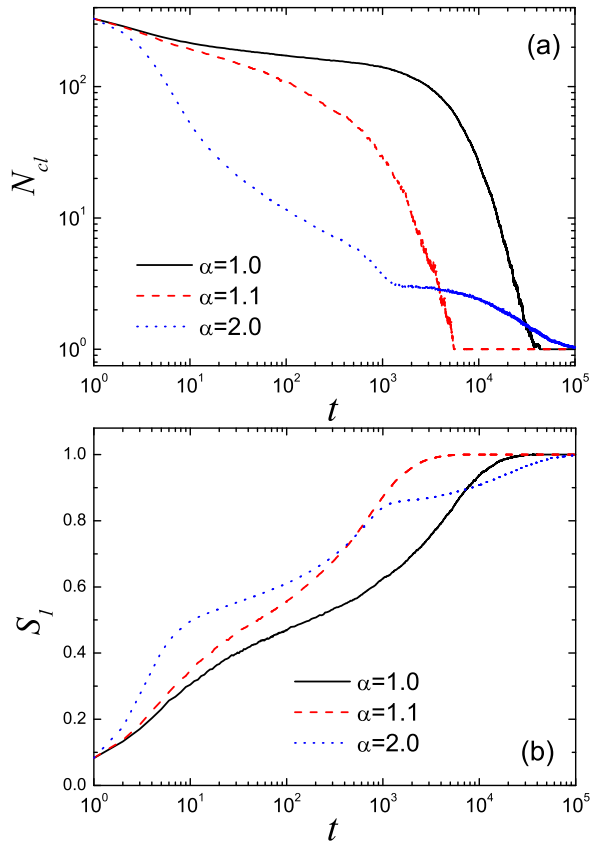}} \caption{(Color
online.) (a) The number of opinion clusters $N_{cl}$ and (b) the
normalized size of the largest cluster $S_{1}$ as a function of time
$t$ for different values of $\alpha$ on a $50\times50$ square
lattice. Each data point is obtained by averaging over 2000
different realizations.} \label{cluster}
\end{center}
\end{figure}

\begin{figure}
\begin{center}
 \scalebox{0.7}[0.7]{\includegraphics{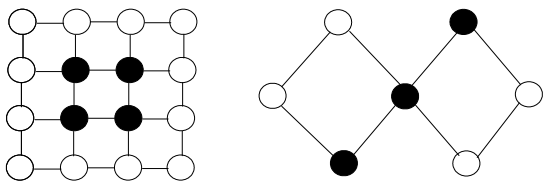}}
\caption{Illustration of the configuration for the coexistence of
two opinions for $\alpha\rightarrow\infty$.} \label{map}
\end{center}
\end{figure}

\begin{figure}
\begin{center}
\scalebox{0.45}[0.45]{\includegraphics{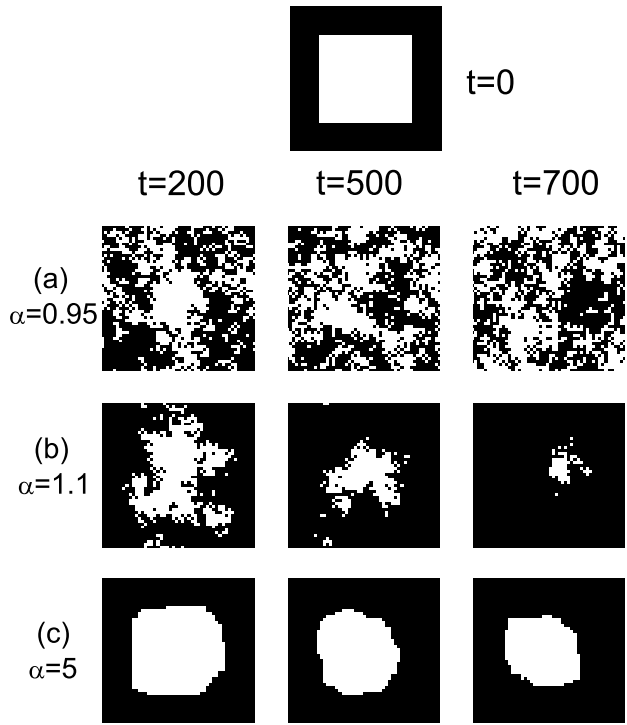}}
\caption{Snapshots of opinion patterns on a $50\times50$ square
lattice. Initially ($t=0$), we set a subregion of $30\times30$ at
the center where all agents have the opinion $+1$ (white), but the
other nodes carry the opinion $-1$ (black). Initially the fraction
of the opinion +1 is $\rho_{+}(0)=9/25=0.36$. (a) $\alpha=0.95$,
$\rho_{+}(200)=0.41$, $\rho_{+}(500)=0.4576$, $\rho_{+}(700)=0.474$;
(b) $\alpha=1.1$, $\rho_{+}(200)=0.2712$, $\rho_{+}(500)=0.1276$,
$\rho_{+}(700)=0.0278$; (c) $\alpha=5$, $\rho_{+}(200)=0.3232$,
$\rho_{+}(500)=0.2428$, $\rho_{+}(700)=0.1916$.} \label{patt}
\end{center}
\end{figure}

To understand the process of convergence to consensus, we study the
evolution of opinion clusters. A opinion cluster is a connected
component (subgraph) fully occupied by nodes holding the same
opinion. Figure~\ref{cluster} shows the number of opinion clusters
$N_{cl}$ and the normalized size of the largest cluster $S_{1}
=S_{max}/N$ as a function of $t$ for different values of $\alpha$,
where $S_{max}$ is the size of the largest cluster. We see that
eventually $N_{cl}$ decreases to 1 and $S_{1}$ increases to 1 for
large $t$. For small values of $\alpha$, agents do not select the
local majority opinion with large probability, thus it becomes
difficult for them to form large opinion clusters. For with
$\alpha=1.0$, at the beginning, $N_{cl}$ decreases and $S_{1}$
increases much more slowly. For $\alpha=2.0$, initially $N_{cl}$
decreases and $S_{1}$ increases more quickly than the cases of
$\alpha=1.0$ and $\alpha=1.1$. However, when there are only two or
three opinion clusters remained in the system, $N_{cl}$ decreases
and $S_{1}$ increases very slowly for $\alpha=2.0$, indicating that
the merging of different clusters becomes difficult for large values
of $\alpha$. In particular, for $\alpha\rightarrow\infty$, the
system cannot reach consensus in some cases. As shown in
Fig.~\ref{map}, two different opinions can coexist forever when
$\alpha\rightarrow\infty$. For moderate values of $\alpha$
($\alpha=1.1$), large opinion clusters can be formed more rapidly
than for low values of $\alpha$ ($\alpha=1.0$), and the their
emergence is facilitated than the cases of larger values of $\alpha$
(e.g., $\alpha=2.0$) as well. Thus the convergence time becomes
minimum for some moderate value of $\alpha$. It can also be noted
that the evolution of $S_{1}$ for different values of $\alpha$ is
similar to that of $\eta$ (see Fig.~\ref{order}).

To assess how different values of $\alpha$ affect the evolution of
the opinion clusters, we design a numerical procedure to investigate
the evolution of spatial patterns of opinion. Specifically, in the
$50\times50$ square lattice, initially we set a subregion of
$30\times30$ at the center where all agents have the opinion $+1$,
but the other nodes carry the opinion $-1$, as shown in
Fig.~\ref{patt}. We see that for $\alpha <1$, e.g., $\alpha = 0.99$,
the boundary between the two opinion clusters gradually become
blurred and they tend to be well-mixed with approximately equal
densities. In this case, global consensus is less likely due to the
difficulty to form opinion clusters. For moderate values of
$\alpha$, e.g., $\alpha = 1.1$, the boundaries become irregular with
time but finally the $+1$ cluster vanishes. For high values of
$\alpha$, e.g., $\alpha = 5.0$, the boundaries are clear and the
inside cluster shrinks but at a speed lower than that for
$\alpha=1.1$.

\begin{figure}
\begin{center}
\scalebox{0.4}[0.4]{\includegraphics{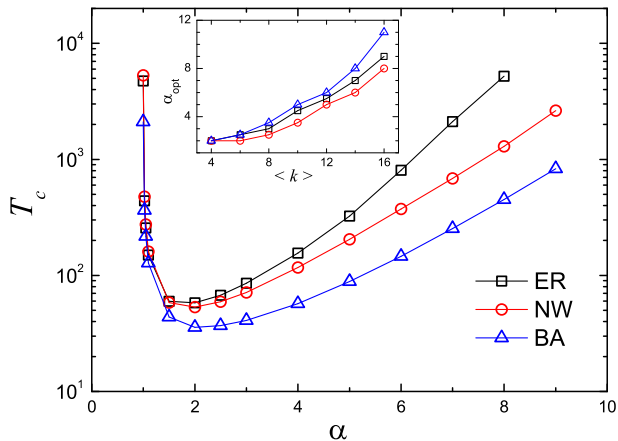}} \caption{(Color
online.) Convergent time $T_{c}$ as a function of $\alpha$ for ER
random graphs, NW small-world networks and BA scale-free networks.
The average connectivities of ER, NW and BA networks are 4. The
inset shows the value of $\alpha_{opt}$ as a function of the average
connectivity $\langle k \rangle$ for the three types of networks.
The network size is $N=3000$. Each data point is obtained by
averaging over 100 different network realizations with 20 runs for
each realization.} \label{network}
\end{center}
\end{figure}

The existence of an optimal value of $\alpha$ for which global
consensus can be achieved rapidly is not restricted to the
square-lattice structure. In fact, we have observed a similar
behavior for complex networks including Erd\"{o}s-R\'{e}nyi random
graphs (ER) \cite{ER}, Newman-Watts small-world networks (NW)
\cite{NW} and Barab\'{a}si-Albert scale-free networks (BA)
\cite{BA}. We also observe that, the optimal value of $\alpha$ tends
to increase with the average connectivity $\langle k \rangle$ (see
the inset of Fig.~\ref{network}). From Fig.~\ref{network}, one can
see that the topological structure affects the convergent time
$T_{c}$. Among the three complex networks, $T_{c}$ for scale-free
networks is shortest when $\langle k \rangle$ and $\alpha$ is fixed.

\section{Conclusion}

In summary, we have proposed a nonlinear voter model to study the
convergence to global consensus on lattices and complex networks. An
agent $i$ selects one of binary opinions with the probability that
is proportional to the power of the number of agents carrying this
opinion among agent $i$ and its nearest neighbors. The power
exponent $\alpha$ is introduced to govern selective probability. It
is found that there exists an optimal value of $\alpha$ leading to
the shortest convergent time. We have explained such phenomenon in
terms of the evolution of the opinion clusters. For too small values
of $\alpha$, the formation of big opinion clusters is slow. For very
high values of $\alpha$, the merging of different opinion clusters
becomes difficult. Taken together, the shortest convergent time can
be realized at moderate values of $\alpha$. Our results indicate
that, the consensus would be quickly reached if agents suitably
follow the local majority opinion.

\begin{acknowledgments}
HXY and BHW were supported by the National Basic Research Program of
China (973 Program No. 2006CB705500); the National Important
Research Project (Grant No. 91024026); the National Natural Science
Foundation of China (Grant Nos. 11005001, 10975126 and 10635040),
the Specialized Research Fund for the Doctoral Program of Higher
Education of China (Grant No. 20093402110032). WXW and YCL were
supported by AFOSR under Grant No. FA9550-10-1-0083.
\end{acknowledgments}


\begin{references}
\bibitem{note} We have checked that the qualitative results do not change when asynchronous
update is used.
\end{references}
\end{document}